\documentclass[journal,10pt, one column]{IEEEtran}
\usepackage{setspace}
\usepackage[ruled]{algorithm2e}
\usepackage{algorithmic}
\usepackage{amsmath}
\usepackage{amsthm}
\usepackage{float}
\usepackage{amsfonts}
\usepackage{amssymb}

\usepackage{mathrsfs}
\usepackage{graphicx}
\usepackage{subfigure}
\usepackage{amsthm}
\usepackage[noadjust]{cite}

\usepackage{tabularx}
\usepackage{multirow}
\usepackage{adjustbox}
\usepackage{tablefootnote}
\usepackage{threeparttable}

\newcommand{\norm}[1]{\left|\left|#1\right|\right|}

\begin{document}

\title{Proportionate Adaptive Filtering under Correntropy Criterion in Impulsive Noise Environments}

\author{Vinay Chakravarthi Gogineni$^1$, {Subrahmanyam Mula$^2$}\\
Department of Electronics and Electrical Communication Engineering\\
Indian Institute of Technology, Kharagpur, INDIA\\
E.Mail : $^1\;$vinaychakravarthi@ece.iitkgp.ernet.in, $^2\;$svmula@iitkgp.ac.in}

\maketitle

\begin{abstract}
An improved proportionate adaptive filter based on
the Maximum Correntropy Criterion (IP-MCC) is proposed for identifying the system with variable sparsity in an impulsive noise environment. Utilization of MCC mitigates the effect of impulse noise while the improved proportionate concepts exploit the underlying system sparsity to improve the convergence rate. Performance analysis of the proposed IP-MCC is carried out in the steady state and our analysis reveals that the steady state Excess Mean Square Error (EMSE) of the proposed IP-MCC filter is similar to the MCC filter. The proposed IP-MCC algorithm outperforms the state of the art algorithms and requires much less computational effort. The claims made are validated through exhaustive simulation studies using the correlated input.
\end{abstract}

\begin{IEEEkeywords}
Impulsive Interference, Maximum Correntropy Criterion, Sparse Adaptive Filters, Proportionate-type Adaptive Filters.
\end{IEEEkeywords}

\IEEEpeerreviewmaketitle

\section{Introduction}
The popular Least Mean Square (LMS) family of algorithms that developed
under Minimum Mean Square Error (MMSE) criterion (i.e., minimizing the $\ell_{2}$-norm of the error) are effective in Gaussian noise environment, however, perform poorly for the non-Gaussian impulsive interference, such as low frequency atmospheric noise, many types of man-made noise and underwater acoustic noise \cite{implusenoise}. Impulsive interference is generally characterized by a heavy-tail distribution, and research shows that a lower-order statistical measure of
the error in cost function offers more robustness against impulsive noise \cite{APSA}. The Sign Algorithm \cite{SA} which was later extended to  the Normalized Sign Algorithm (NSA) \cite{NSA}, employ the mean absolute value of the error (i.e., $\ell_{1}$- norm of the error) as the cost function, exhibit robustness against the impulsive interference. To attain the improved convergence rate in colored input conditions, the Affine Projection Sign Algorithm (APSA) \cite{APSA} is proposed by minimizing $\ell_{1}$ norm of the \emph{a postiriori} error vector.

For applications like Network Echo Cancellation (NEC), the system  (network echo path) to be estimated is sparse in nature (i.e., impulse response contains very few active coefficients while the rest of the coefficients magnitude is close to zero) \cite{NEC}. Since the APSA is sparsity agnostic, it is not a best choice for sparse system identification. Inspired from the proportionate adaptation \cite{PNLMS}, two proportionate affine projection sign algorithms, namely real-coefficient Proportionate APSA (RP-APSA) \cite{RIPAPSA} and real-coefficient Improved Proportionate APSA (RIP-APSA) \cite{RIPAPSA} are proposed by employing the proportionate concepts \cite{PNLMS} and improved proportionate concepts \cite{IPNLMS} to the APSA. Both the RP-APSA and RIP-APSA yield the faster convergence rate and lower misadjustmet over APSA at the cost of increased complexity. To reduce the computational complexity, the RIP-APSA was then modified to the memory Improved Proportionate APSA (MIP-APSA) \cite{MIPAPSA}. Recently, by employing the Lorentzian norm, a Lorentzian based Adaptive Filter (LAF) is proposed in \cite{LHTAF}, which was later extended to the Normalized Lorentzian based Hard Thresholding Adaptive Filter (Normalized LHTAF) and Normalized Lorentzian based Variable Hard Thresholding Adaptive Filter (Normalized LVHTAF) by incorporating the Iterative Hard Thresholding concepts \cite{IHT} into the Normalized LAF. Since both the Normalized LVHTAF and LHTAF use the hard thresholding operator $H_{K}$ (that sets all coefficients of weight vector to zero, except $K$ largest (in magnitude) coefficients), their performance is hugely dependent on the assumed sparsity value $K$. In general, the sparsity of the network echo path may vary with time and context and hence they sufferers while tracking these variations. Moreover, they involve huge computational complexity, hence not suitable for real-time applications.

To address these issues of Lorentzian based algorithms (i.e., robustness against the variable sparsity and computational complexity), in this paper, we propose an improved Proportionate MCC (IP-MCC) algorithm by combining the recently proposed Maximum Correntropy Criterion (MCC) based adaptive filter \cite{MCCTSP, MCC,MCCMSD, AKWMCC, AKWMCC-Anal, St-MCC, APAMCC, GMCC} with improved proportionate concepts \cite{IPNLMS}. Usage of MCC makes the algorithm robust against the impulsive interference and the improved proportionate adaptation ensures high convergence rate by exploiting the underlying system sparsity. Moreover, improved proportionate concepts also make the algorithm robust against the time varying system sparsity. Using the energy conservation approach \cite{Nonlinearities}, we carried out the performance analysis in steady state and the analysis shows that the steady state EMSE of Pt-MCC is same as that of MCC. The experimental results demonstrate that the proposed algorithm outperforms the state of the art algorithms (in terms of convergence rate and steady state misalignment) with much reduced computational complexity.
\section{Algorithm Design}
We consider here the problem of identifying a system that takes a input
signal $u(n)$ and produces the observable output $d(n)={\bf
u}^{T}(n) {\bf w}_{opt} + \vartheta(n) $, where ${\bf u}(n)=[{u}(n),
{u}(n-1), \cdots, {u}(n-L+1)]^{T}$ is the input data vector at time
index $n$, ${\bf w}_{opt}$ is a $L \times 1$ system impulse response
vector (to be identified) which is known \emph{a priori} to be
sparse with variable sparsity and $\vartheta(n)$ constitutes of the
observation noise plus impulsive interference with mean zero and
variance $\sigma^{2}_{\vartheta}$ which is taken to be i.i.d. and
independent of input $u(m)$ for all $n$, $m$.

The MCC based stochastic gradient adaptive filter in \cite{MCC} is derived by maximizing the cost function $E\big[\text{exp}\big(-\frac{e^{2}(n)}{2 \hspace{1mm}\sigma^{2}}\big) \big]$ and the corresponding update equation is given by 
\begin{equation}\label{MCC}
{\bf w}(n+1)={\bf w}(n)+ \mu \hspace{1mm}
\text{exp}\big(-\frac{e^{2}(n)}{2 \sigma^{2}} \big) \hspace{1mm}
e(n) \hspace{1mm} {\bf u}(n),
\end{equation}
where $\mu$ is the adaptation step size, $e(n)= d(n)- {\bf w}^{T}(n) {\bf u}(n)$ is the estimation error and $\sigma$ is the kernel width. 

The MCC filter in \eqref{MCC} is sparse agnostic which means that it can't exploit the underlying system sparsity. To achieve this, MCC filter coefficients can be proportionately adapted by pre-multiplying the update vector with the proportionate gain matrix
${\bf G}(n)$. The resultant algorithm is termed as the Proportionate MCC  (PMCC) adaptive filter and its weight update is given as follows:
\begin{equation}\label{Pt-MCC}
{\bf w}(n+1)={\bf w}(n)+ \mu \hspace{1mm}
\text{exp}\big(-\frac{e^{2}(n)}{2 \sigma^{2}}\big) \hspace{1mm}
e(n) \hspace{1mm} {\bf G}(n) \hspace{1mm} {\bf u}(n),
\end{equation}
where ${\bf G}(n)=diag\{g_{0}(n), g_{1}(n), \cdots, g_{L-1}(n)\}$,
distributes the adaption energy among the filter taps in
proportional to the individual filter tap magnitude i.e., $g_{i}(n)
\propto |w_{i}(n)|$ \cite{PNLMS}. In general, sparsity of the echo path
varies over time and for effective identification of these time
varying sparse systems, improved Proportionate concepts \cite{IPNLMS} are
preferred. The gain factors, $g_{i}(n)$ of these improved
proportionate concepts \cite{IPNLMS} are given by,
\begin{equation}\label{gIPNLMS}
g_{i}(n)= \frac{1- \alpha}{2 \hspace{1mm} L}  +  (1+ \alpha)
\hspace{1mm} \frac{ |w_{i}(n)|} {2 \hspace{0.5mm} \|{\bf w}(n)\|_{1}
+ \epsilon_{p}},
\end{equation}
where $-1 \leq \alpha \leq 1$ and $\epsilon_{p}$ is a small positive
constant, employed to avoid division by zero.

Please note that the weighted
Euclidian norm of the input (i.e., ${\bf u}^{T}(n) {\bf G} (n) {\bf
u}(n) $) which is generally present in PNLMS algorithm is omitted
in the PMCC \eqref{Pt-MCC}. As it is proved in \cite{PLMS} that the penalty for omission of this normalization is negligible and it saves huge computational burden. Since we omitted the normalization term in \eqref{Pt-MCC}, the gain factors given in \eqref{gIPNLMS} can not be applied directly to the proposed Pt-MCC algorithm. For this, the gain factors are reformulated to the following:
\begin{equation}\label{gIPLMS}
g_{i}(n)= \frac{1- \alpha}{2}  +  (1+ \alpha) \hspace{1mm} \frac{L
\hspace{1mm} |w_{i}(n)|} {2 \hspace{0.5mm} \|{\bf w}(n)\|_{1}  +
\epsilon_{p}}.
\end{equation}
the resultant algorithm is named as improved Proportionate MCC
(IP-MCC). The proposed IP-MCC is summarized in Algorithm $1$. 
\begin{algorithm}
\DontPrintSemicolon 
Initialization : $ {\bf w}(0)= {\bf 0}$, \;
Parameters : $\mu, \sigma, \alpha, \epsilon_{p}$ \;
\vspace{2mm}
Updation :\;
 \vspace{1mm}
 $e(n) = d(n)- \textbf{w}^{T}(n) \textbf{u}(n)$\; 
 $g_{i}(n)= \frac{1- \alpha}{2}  +  (1+ \alpha) \hspace{1mm} \frac{L
\hspace{1mm} |w_{i}(n)|} {2 \hspace{0.5mm} \|{\bf w}(n)\|_{1}  +
\epsilon_{p}},\hspace{1mm} \text{for} \hspace{1mm} i=0, \cdots, L-1$\;
 $\textbf{G}(n)$ =
$diag(g_{0}(n),g_{1}(n),...g_{L-1}(n))$ \;
$\triangle {\bf w}(n) = \text{exp}\big(-\frac{e^{2}(n)}{2 \sigma^{2}}\big) \hspace{1mm}
{\bf G}(n) \hspace{1mm} {\bf u}(n)$\;
$\textbf{w}(n+1)=\textbf{w}(n)+ \mu \hspace{1mm} \triangle {\bf w}(n)$\;
\caption{Proposed IP-MCC Algorithm} 
\label{IP-MCC}
\end{algorithm}

\section{Steady State Performance Analysis}
The proposed IP-MCC can be seen as the transform domain MCC with transformed input ${\bf s}(n)= {\bf G}^{\frac{1}{2}}(n) \hspace{1mm} {\bf u}(n)$ and transformed filter coefficient vector ${\bf w}_{t}(n)= {\bf G}^{\frac{-1}{2}}(n) \hspace{1mm} {\bf w}(n)$, as suggested in \cite{PtNLMS} in the context of PNLMS analysis. Defining the weight deviation vector of the proposed IP-MCC $\widetilde{{\bf w}}(n)= {\bf w}_{opt} - {\bf w}(n)$ and its corresponding transformed weight deviation vector $\widetilde{{\bf w}}_{t}(n)= {\bf G}^{\frac{-1}{2}}(n) \hspace{1mm} \widetilde{{\bf w}}(n) = {\bf G}^{\frac{-1}{2}}(n) \hspace{1mm} {\bf w}_{opt} - {\bf w}_{t}(n)$, it is easy to check that $e(n)= \widetilde{{\bf w}}^{T}(n) \hspace{1mm} {\bf u}(n) + \vartheta(n)= \widetilde{{\bf w}}^{T}_{t}(n) \hspace{1mm} {\bf s}(n) + \vartheta(n) = e_{a}(n) + \vartheta(n)$, where $e_{a}(n)= \widetilde{{\bf w}}^{T}_{t}(n) \hspace{1mm} {\bf s}(n)$ is the \emph{a priori} error. As in \cite{PtNLMS}, assuming ${\bf G}^{\frac{-1}{2}}(n) \hspace{1mm} \widetilde{{\bf w}}(n+1) ={\bf G}^{\frac{-1}{2}}(n+1) \hspace{1mm} \widetilde{{\bf w}}(n+1)$ (which is reasonable assumption particularly near convergence), from \eqref{Pt-MCC}, we then can write, 
\begin{equation}\label{Wd_Pt-MCC}
\widetilde{{\bf w}}_{t}(n+1)= \widetilde{{\bf w}}_{t}(n) - \mu \hspace{1mm}
f\big(e(n)\big)  \hspace{1mm} {\bf s}(n),
\vspace{-2mm}
\end{equation} 
where $f\big(e(n)\big) = \text{exp}\big(-\frac{e^{2}(n)}{2 \sigma^{2}}\big) \hspace{1mm} e(n)$. Using the energy conservation approach \cite{Nonlinearities}, we then can have the following relation:
\begin{equation}\label{MSWd_Pt-MCC}
\begin{split}
E \Big[\|\widetilde{{\bf w}}_{t}(n+1)\|^{2}\Big]&= E \Big[\|\widetilde{{\bf w}}_{t}(n)\|^{2}\Big] - 2\mu \hspace{1mm} E\Big[ e_{a}(n) \hspace{1mm} f\big(e(n)\big) \Big] \\
&\hspace{2mm}+ \mu^{2} E\Big[ \| {\bf s}(n)\|^{2} \hspace{1mm} f^{2}\big(e(n)\big)\Big].
\end{split}\raisetag{0.8\baselineskip}
\end{equation}
To evaluate the EMSE, we assume the following (which are commonly used in the analysis of adaptive filters with error non-linearities \cite{Nonlinearities}):
\begin{itemize}
\item[] {\bf A}$1$: The noise signal $\vartheta(n)$ is zero mean i.i.d. with variance $\sigma_{\vartheta}^{2}$ and taken to be independent of input $u(m)$ for all $n$ and $m$.

\item[] {\bf A}$2$: The \emph{a priori} error $e_{a}(n)$ is zero mean and independent of the noise $\vartheta(n)$.

\item[] {\bf A}$3$: The filter is long enough such that $e_{a}(n)$ is Gaussian, and $\|{\bf s}(n) \|^{2} $ is asymptotically uncorrelated with $f^{2}\big(e(n)\big)$, implying near convergence 
$E\left[ \| {\bf s}(n)\|^{2} \hspace{1mm} f^{2}\big(e(n)\big)\right]= Trace\big( {\bf S}\big) \hspace{1mm} E\left[ f^{2}\big(e(n)\big)\right]$, where ${\bf S}= E[ {\bf s}(n) \hspace{1mm} {\bf s}^{T}(n)]=E\big[{\bf G}^{\frac{1}{2}}(n) \hspace{1mm} {\bf u}(n) \hspace{1mm} {\bf u}^{T}(n) \hspace{1mm} {\bf G}^{\frac{1}{2}}(n) \big]$. As ${\bf G}(n)$ varies slowly with time when compared with ${\bf w}(n)$, we can assume ${\bf G}(n)$ is independent of ${\bf u}(n)$ \cite{PtNLMS}, then we can write, ${\bf S}=E\big[{\bf G}^{\frac{1}{2}}(n) \hspace{1mm} E[ {\bf u}(n) \hspace{1mm} {\bf u}^{T}(n)] \hspace{1mm} {\bf G}^{\frac{1}{2}}(n) \big]= E\big[{\bf G}^{\frac{1}{2}} (n) \hspace{1mm} {\bf R} \hspace{1mm}{\bf G}^{\frac{1}{2}}(n) \big] $.
\end{itemize}
In order to obtain the steady state EMSE, we now assume the time index $n$ to belong to the steady state of the IP-MCC filter (i.e., $n$ is sufficiently large). One can then assume IP-MCC is stable and $E \big[\|\widetilde{{\bf w}}_{t}(n)\|^{2}\big]$ has converged to its steady state value $\lim\limits_{n \to \infty} E \big[\|\widetilde{{\bf w}}_{t}(n)\|^{2} \big]$. Therefore, from \eqref{MSWd_Pt-MCC}, in steady state, we then have

\begin{equation}\label{MSWd_Pt-MCC1}
\begin{split}
\lim\limits_{n \to \infty} E\Big[ e_{a}(n) \hspace{1mm} f\big(e(n)\big) \Big] = \frac{\mu}{2} \hspace{1mm} Trace({\bf S}) \hspace{1mm}\lim\limits_{n \to \infty} E\Big[f^{2}\big(e(n)\big)\Big].
\end{split}
\end{equation}
The above result is similar to the one presented in \cite{MCCMSD}. Following the same procedure given in \cite{MCCMSD}, the steady state EMSE of the proposed IP-MCC: $\xi= \lim\limits_{n \to \infty} E[e^{2}_{a}(n)]$ can be obtained as, \newline
{\bf Gaussian Noise}:
\begin{equation}\label{Gaussian_EMSE}
\begin{split}
\xi= \frac{\mu}{2} \hspace{2mm} Trace( {\bf S}) \hspace{2mm} \frac{(\xi+ \sigma^{2}_{\vartheta}) \hspace{1mm} (\xi+ \sigma^{2}_{\vartheta} + \sigma^{2})^{\frac{3}{2}}} {(2\xi+ 2\sigma^{2}_{\vartheta} + \sigma^{2})^{\frac{3}{2}}}.
\end{split}
\end{equation}
{\bf Impulsive Noise}:
\begin{equation}\label{Impulsive_EMSE}
\begin{split}
\xi= \frac{\mu}{2} \hspace{1mm} Trace({\bf S}) \frac{E\Big[exp(- \frac{\vartheta^{2}}{\sigma^{2}}) \hspace{1mm} \vartheta^{2}\Big]} {E\Big[exp(- \frac{\vartheta^{2}}{2\sigma^{2}}) \hspace{1mm} (1- \frac{\vartheta^{2}}{\sigma^{2}})\Big]}.
\end{split}
\end{equation}
For a white input ${\bf u}(n)$ with variance $\sigma_{u}^{2}$, the autocorrelation matrix ${\bf R}= \sigma_{u}^{2} \hspace{1mm} {\bf I}$, thus, we have  ${\bf S}= \sigma^{2} E[{\bf G}(n)]$ and $Trace({\bf S})=\sigma^{2}_{u} \hspace{1mm} Trace(E[{\bf G}(n)])=  \sigma^{2}_{u} \sum\limits_{i=0}^{L-1} E[g_{i}(n)]= L \sigma^{2}_{u}$. Substituting this in \eqref{Gaussian_EMSE} and \eqref{Impulsive_EMSE}, it is easy to verify that the steady state EMSE of the proposed IP-MCC is same as that of MCC (which is worked out in \cite{MCCMSD}). Therefore, if we incorporate proportionate concepts into MCC, we can achieve better convergence rate without compromising the steady state EMSE performance.
\section{Computational Complexity}
The computational complexity of the proposed IP-MCC is compared with
the MIP-APSA \cite{MIPAPSA} and the Lorentzian based algorithms \cite{LHTAF} in terms of basic arithmetic operations such as, additions, multiplications, divisions, etc. With filter length $L$ and projection order $P$ (for data reuse algorithms), the corresponding complexities are shown in Table~\ref{Table1}.

\begin{table}[h!]
\centering
\caption{Comparison of Computational Complexity}
\label{Table1}
\begin{threeparttable}
\begin{tabular}{ |c|m{4.3em}|m{4.5em}|m{2.2em}|m{2.5em}| m{2.5em}| }
 \hline
Algorithm  & Add & Mul & Div  & Sqrt$/$ Exp & Sort$/$ Median\\ [1ex]

  \hline
  \hline
 {MIP-APSA \cite{MIPAPSA}}  & $L(2P + 3) $ & \shortstack{$L(P+4)$\\$+ 2$} & $2$ & $1$ & $0$ \\  
  \hline 
  {\shortstack{Normalized \\LHTAF \cite{LHTAF}}}  & \shortstack{$L(3P +1)$\\$+ (P-1)$} & \shortstack{$L(3P +2)$\\$+ 4(P+1)$} & $P+1$ & $0$ & $2$ \\  
  \hline
   {\shortstack{Normalized \\LVHTAF \cite{LHTAF}}}  & \shortstack{$L(3P +1)$\\$+ (P+1)$} & \shortstack{$L(3P +2)$\\$+ 4(P+2)$} & $P+1$ & $2$ & $2$ \\  
  \hline
  {\shortstack{IP-MCC \\ $[$Proposed$]$ } }  & \shortstack{$4L$} & \shortstack{$4L +5$} & $1$ & $1$ & $0$ \\  
  \hline
\end{tabular}
\begin{tablenotes}
      \item Add: Additions, Mult: Multiplications, Div: Divisions, Sqrt: Square root, Exp: Exponential, Sort: Sorting.  
    \end{tablenotes}
\end{threeparttable}
\end{table}

From Table~\ref{Table1}, it can be noticed that the proposed IP-MCC requires $L(2P -1)$
fewer additions and $LP-3$ fewer multiplications than the MIP-APSA algorithm. Moreover, the MIP-APSA also requires $L (P-1)$ extra memory locations to store ${\bf G}(n-k) \hspace{1mm} {\bf u}(n-k)$, $k=1, 2, \cdots, P-1$. Similarly, 
when compared to the Normalized LVHTAF algorithm, the proposed IP-MCC requires $3L(P-1) +(P+1)$ fewer additions and $L(3P -2) + (4P +3) $ fewer multiplications. Moreover, Normalized LVHTAF also requires $P$ excess divisions, $1$ square root and $2$ sorting operations. One can see that the computational efficiency of proposed IP-MCC increases with $L$ and $P$.  

To quantify the achieved improvement in computational efficiency, we consider a typical case of $L=512$ and $P=10$ (which is considered for simulations) and the corresponding computational gain of IP-MCC is given below:
\begin{itemize}
\item  When compared to the MIP-APSA algorithm, the proposed IP-MCC algorithm requires $82.6\%$ fewer additions and $71.43\%$ fewer multiplications. Furthermore, MIP-APSA also requires $4608$ more memory locations.

\item When compared to the Normalized LVHTAF algorithm, the proposed IP-MCC algorithm requires $87.1\%$ fewer additions and $87.5\%$ fewer multiplications. Moreover, Normalized LVHTAF also requires $10$ extra divisions, $1$ square root and $2$ sorting operations of vector lengths $512$ (filter length) and $10$ (projection order).
\end{itemize}
From the above discussion, it is evident that the proposed IP-MCC algorithm  significantly improves computational efficiency, there by making it feasible for real-time implementations. In the following section through detailed simulations, we demonstrate that even with this reduced complexity, the proposed IP-MCC algorithm achieves similar or even better performance (in terms of convergence rate and Mean Square Deviation (MSD)) over the MIP-APSA and Normalized LVHTAF/LHTAF algorithms.
\section{Simulation Studies and Discussion}
\begin{figure}[h!]
\centering
\includegraphics[height=25mm,width=80mm]{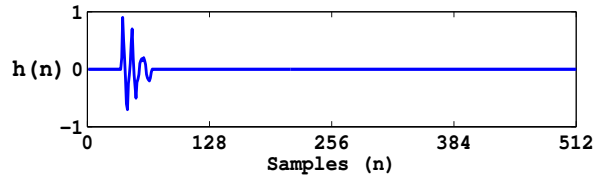}
\caption{Sparse system impulse response of length $L=512$ with $32$ active taps.} \label{H1_response}
\end{figure}

In this section, we provide the comparative learning curves (MSD: $\|{\bf w}_{opt} - {\bf w}(n)\|^{2}$ in dB vs  iteration index $n$) to compare the performances of proposed IP-MCC, MIP-APSA and Normalized LVHTAF/LHTAF. The simulation results are obtained by averaging over $5000$ independent runs.
A series of simulations are conducted in system identification context using colored input. The unity variance colored Gaussian input $u(n)$ is generated by filtering a zero-mean, unit variance, white Gaussian signal through a first-order system $G(z)=\frac{\sqrt{1- \theta^{2}}}{1- \theta z^{-1}}$ ($\theta$ was taken to be $0.9$ for simulations). We considered the  additive noise $\vartheta(n)$ as a mixture of  Gaussian white noise $s(n)\sim \mathcal{N}(0, \sigma^{2}_{s})$ and impulsive noise $q(n)$, i.e., $\vartheta(n) = s(n) +  q(n)$. The interference signal is generated as the product of a Bernoulli process and a Gaussian process \cite{APSA}, i.e., $q(n)= B(n) I(n)$, where $B(n)$ is a Bernoulli process with probability of occurrence $Pr(B(n)=1)=p$ and $I(n) \sim \mathcal{N}(0, \sigma^{2}_{I})$, with $\sigma^{2}_{I} >> \sigma^{2}_{s}$. For simulations, we considered $\sigma^{2}_{s}=0.01$ and $\sigma^{2}_{I}=1000$. For all the experiments, the Projection order is fixed at $P=10$ for both MIP-APSA and Normalized LHTAF/LVHTAF algorithms. The parameter $\alpha$ in \eqref{gIPNLMS} and \eqref{gIPLMS} is set to $0$ and the kernel width parameter $\sigma$ in \eqref{Pt-MCC} is fixed at $1.25$. The step sizes are adjusted to attain the same steady state MSD and the corresponding convergence rates are compared. \\ \\
{\emph Experiment} $1$: \newline
In this experiment, we consider a system impulse response ${\bf w}_{opt}$ shown in Fig.~\ref{H1_response}, which is of length $L=512$ with $32$ number of active taps. The probability of occurrence of Bernoulli process is set to $Pr(B(n)=1)=p=0.001$. Please note that this experimental setup (system impulse response and probability of occurance of the impulse noise) is same as in \cite{LHTAF} for fair comparison. 

\begin{figure}[h!]
\centering
\includegraphics[height=65mm,width=85mm]{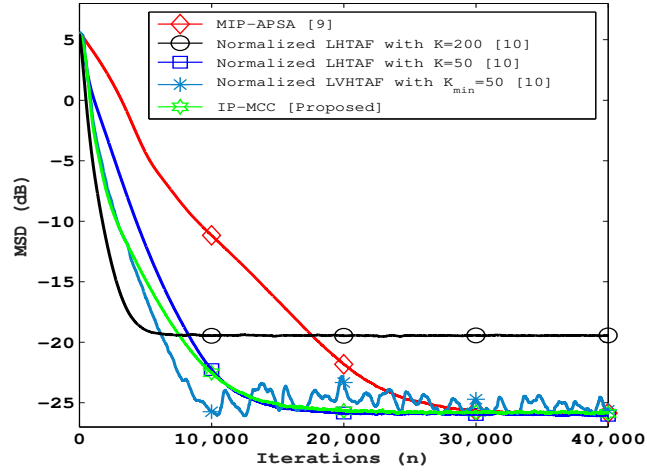}
\caption{Learning curves of the Proposed IP-MCC vis-a-vis those of a few recently proposed algorithms to identify the system shown in Fig.~\ref{H1_response} with $\mu_{MIP-APSA}= 0.001$, $\mu_{Normalized \hspace{1mm} LVHTAF/LHTAF}=0.9$,
$\mu_{IP-MCC}=0.00097$ for $Pr(B(n)=1)=p=0.001$} \label{Elsevier-1}
\end{figure}

\begin{figure}[h!]
\centering
\includegraphics[height=65mm,width=85mm]{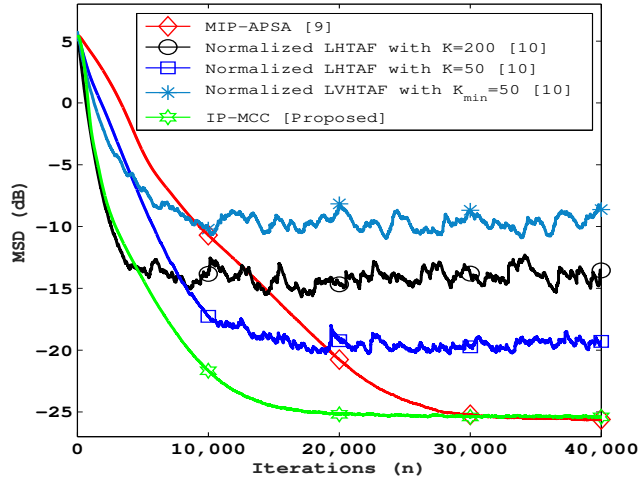}
\caption{Learning curves of the Proposed IP-MCC vis-a-vis those of a few recently proposed algorithms to identify the system shown in Fig.~\ref{H1_response} with $\mu_{MIP-APSA}= 0.001$, $\mu_{Normalized \hspace{1mm} LVHTAF/LHTAF}=0.9$,
$\mu_{IP-MCC}=0.00097$ for $Pr(B(n)=1)=p=0.001$ for $Pr(B(n)=1)=p=0.05$} \label{Elsevier-2}
\end{figure}

For Normalized LHTAF, we considered two values of assumed sparsity $K= 50, 200$, where the actual sparsity of the system is $32$. Please note that in practical scenarios the actual sparsity is unknown and thus, $K=50$ is an optimistic choice where as $K=200$ is little pessimistic. The learning curves are plotted in Fig.~\ref{Elsevier-1}, and the results of MIP-APSA and Normalized LVHTAF/LHTAF are similar to the ones presented in \cite{LHTAF}. The proposed IP-MCC outperforms the MIP-APSA and Normliaed LHTAF, but exhibits slightly poor convergence rate than the Normalized LVHTAF. Since the probability of occurrence of Bernoulli process $Pr(B(n)=1)=p=0.001$ (i.e., $1$ in $1000$ samples) is very low, in the next case, we considered a more realistic scenario with $Pr(B(n)=1)=p=0.05$ (i.e., $5$ in $100$ samples) as suggested in \cite{LLAD} with all other parameters remaining same. Fig.~\ref{Elsevier-2} shows the learning curves for this case and it is clearly evident that the proposed IP-MCC outperforms all the algorithms significantly. From Fig.~\ref{Elsevier-1} and Fig.~\ref{Elsevier-2}, it can also be observed that the performance of the proposed IP-MCC and MIP-APSA remain same, whereas the performance of the Lorentzian based algorithms \cite{LHTAF} degrades drastically when the impulsive noise probability increases.

\begin{figure}[h!]
\centering \subfigure[Higly sparse system with $S_m = 0.8637$\label{Sparse}]{
\includegraphics[height=21mm,width=80mm]{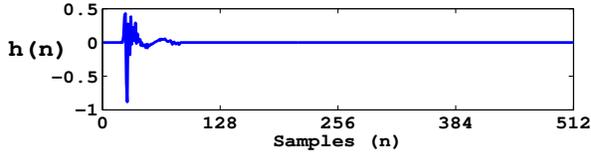} } \hspace{-3mm}
\subfigure[Semi sparse system with $S_m = 0.5560$\label{Dispersive}]{
\includegraphics[height=21mm,width=80mm]{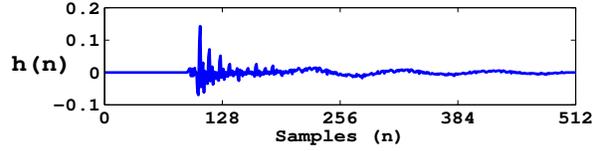} }
\caption{System impulse responses used to evaluate the tracking performance.} \label{System_Response}
\end{figure}

%
{\emph Experiment} $2$: \newline
This experiment (i). evaluates the tracking performance of the proposed IP-MCC algorithm (ii). investigates the robustness of the proposed IP-MCC against the time varying system sparsity. For this, we considered a system impulse response having duration of $0.064$ sec (or equivalently, length $L=512$ taps at $8$ kHz sampling rate) with time varying sparsity. Initially, the system
is taken to be highly sparse with the impulse response shown in Fig.~\ref{Sparse}, for which $S_m = 0.8637$ (where $S_{m}=\frac{L}{L- \sqrt{L}} \left(1- \frac{\norm{{\bf w}_{opt}}_{1}}{\sqrt{L} \hspace{0.3mm} \norm{{\bf w}_{opt}}_{2}}\right)$ \cite{NEC}). After half of the time samples, the
system is changed to a measured acoustic echo
path \cite{NEC} shown in Fig.~\ref{Dispersive}, which is semi sparse with the associated sparseness measure $S_m = 0.5560$.

\begin{figure}[h!]
\centering \subfigure[for $Pr(B(n)=1)=p=0.001$ \label{T-1} ]{
\includegraphics[height=55mm,width=85mm]{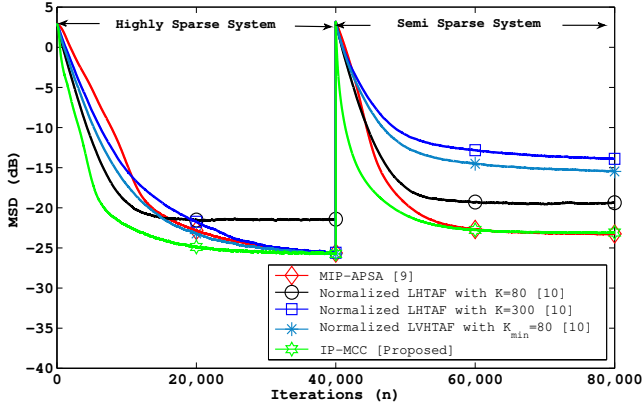} } 
\subfigure[for $Pr(B(n)=1)=p=0.05$ \label{T-2}]{
\includegraphics[height=55mm,width=85mm]{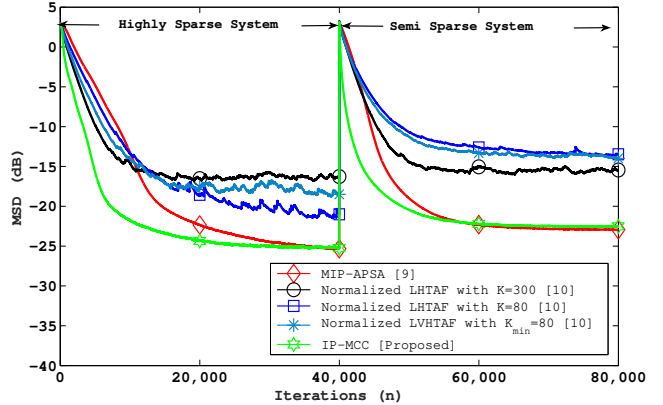} }
\caption{Tracking performance of the Proposed IP-MCC vis-a-vis those of a few recently proposed algorithms for the systems shown in Fig.~\ref{System_Response} with $\mu_{MIP-APSA}= 0.001$, $\mu_{Normalized \hspace{1mm} LVHTAF/LHTAF}=0.35$, $\mu_{IP-MCC}=0.00097$.} \label{System_Response}
\end{figure}

%
%
%
Same as the above, we consider here two cases, i.e., one with low probability of impulsive noise other with high probability of impulsive noise. Figures \ref{T-1} and \ref{T-2} provide the learning curves for these two cases. As shown in Fig.~\ref{T-1}, even under low impulsive noise probability, the Lorentzian based algorithms \cite{LHTAF} fail to track the system when the sprasity of system changes with time. This is intutive as they heavily relay on assumed sparsity value $K$, which is difficult to tune under varaibale sparse conditions. Please note that the tracking experimnet in \cite{LHTAF} had not considered the variable sparsity condition, they simply right circular shifted the impulse response by $30$ positions, which will not change the sparsity of the system. From Fig~\ref{T-2}, we can also observe that the performance of the Lorentzian based algorithms degrades further when the impulsive noise probability is increased to $0.05$. 

From the above two experiments, we can conclude that the proposed IP-MCC is more robust against the impulsive noise as well as time varying system sparsity compared to the existing state of the art algorithms with much reduced complexity. 
\section{Conclusion}
We proposed a novel IP-MCC algorithm for sparse system identification in impulsive noise environments. The proposed algorithm can maintain significantly low computational complexity while achieving similar or even better performance compared with the state of the art algorithms. The proposed algorithm has been evaluated in terms of convergence rate, steady state MSD and tracking performance. Compared to the other algorithms the proposed IP-MCC is more robust against the degree of impulsive noise as well time varying system sparsity. In the mean while, its computational complexity remains low and hence it can be more attractive for practical network echo cancellers.


\bibliographystyle{IEEEtran}

\end{document}